\documentclass[prr,amsmath,amssymb,twocolumn,showpacs,superscriptaddress]{revtex4-2}

\usepackage{graphicx}
\usepackage{subfigure}
\usepackage{adjustbox}
\usepackage{bm}
\usepackage{bbold}
\usepackage{color}
\usepackage{braket}
\usepackage{standalone}
\usepackage{multirow}
\usepackage{tikz}
\usepackage{mathrsfs}
\usepackage{dsfont}
\usepackage[T1]{fontenc}
\usepackage{float}
\usepackage{placeins}
\usepackage[colorlinks,bookmarks=true,citecolor=blue,linkcolor=blue,urlcolor=blue]{hyperref}
\allowdisplaybreaks

\newcommand{\tr}{\textrm{tr}}
\renewcommand{\Re}{\textrm{Re}}
\renewcommand{\Im}{\textrm{Im}}

\begin{document}
\title{Symmetry-induced higher-order exceptional points in two dimensions
}

\author{Anton Montag}
\email{anton.montag@mpl.mpg.de}
\affiliation{Max Planck Institute for the Science of Light, 91058 Erlangen, Germany}
\affiliation{Department of Physics, Friedrich-Alexander-Universit\"at Erlangen-N\"urnberg, 91058 Erlangen, Germany}
\author{Flore K. Kunst}
\email{flore.kunst@mpl.mpg.de}
\affiliation{Max Planck Institute for the Science of Light, 91058 Erlangen, Germany}
\date{\today}

\begin{abstract}
Exceptional points of order $n$ (EP$n$s) appear in non-Hermitian systems as points where the eigenvalues and eigenvectors coalesce.
They emerge if $2(n-1)$ real constraints are imposed, such that EP2s generically appear in two dimensions (2D).
Local symmetries have been shown to reduce this number of constraints.
In this work, we provide a complete characterization of the appearance of symmetry-induced higher-order EPs in 2D parameter space.
We find that besides EP2s only EP3s, EP4s, and EP5s can be stabilized in 2D. Moreover, these higher-order EPs must always appear in pairs with their dispersion determined by the symmetries.
Upon studying the complex spectral structure around these EPs, we find that depending on the symmetry, EP3s are accompanied by EP2 arcs, and two- and three-level open Fermi structures.
Similarly, EP4s and closely related EP5s, which arise due to multiple symmetries, are accompanied by exotic EP arcs and open Fermi structures.
For each case, we provide an explicit example. We also comment on the topological charge of these EPs, and discuss similarities and differences between symmetry-protected higher-order EPs and EP2s.
\end{abstract}

\maketitle

\section{Introduction}

Exceptional points (EPs) are a well-known phenomenon of non-Hermitian (NH) systems~\cite{KatoBook, Heiss2012, Miri2019, Ashida2020}, which in recent years have been extensively studied through the lens of topology~\cite{Bergholtz2021}. EPs are truly NH degeneracies at which not only the eigenvalues but also the eigenvectors coalesce.
To date, most research has focused on EPs of order 2 (EP2s), where the order is set by the number of coalescing eigenvectors.
EP2s appear generically in two-dimensional (2D) parameter space~\cite{Heiss2012}, and they represent the NH analog of nodal points in Weyl semimetals~\cite{Bergholtz2021}. EP2s give rise to unique phenomena, such as the appearance of bulk (i-)Fermi arcs (FAs), where real (imaginary) parts of the eigenenergies of the system coincide~\cite{Berry2004, Kozii2017, Zhou2018}. In higher-dimensional spaces, EP2s are promoted to more complicated structures, such as rings and surfaces~\cite{Xu2017, Cerjan2019, Budich2019, Zhen2015}, which form the boundaries of higher-dimensional (i-)Fermi structures such as 2D Fermi surfaces (FSs)~\cite{Budich2019,Carlstrom2018}.

In order to obtain an $n$th-order EP (EP$n$), the dimension of the parameter space must be larger or equal to the codimension $2n-2$ of the EP$n$~\cite{Delplace2021, Sayyad2022}.
It follows that no EP with $n>2$ can appear naturally in 2D parameter space.
At the same time, it is well established that unitary and anti-unitary symmetries~\cite{Kawabata2019} local in parameter space---namely, parity-time ($\mathcal{PT}$), parity-particle-hole ($\mathcal{CP}$) and pseudo-Hermitian (psH) symmetry, as well as sublattice symmetry (SLS), chiral symmetry (CS) and pseudo-chiral symmetry (psCS)~\cite{Sayyad2022}, cf. Table~\ref{tab:appendix_symmetry}---reduce the number of constraints that need to be imposed in order for EP$n$s to emerge \cite{Budich2019,Yoshida2019,Okugawa2019, Delplace2021, Sayyad2022}, see Table~\ref{tab:symmetries}. 
As such it is possible to induce higher-order EPs in 2D parameter space.
While some theoretical models featuring symmetry-protected EP3s~\cite{Mandal2021,Sayyad2022, Zhang2020, Schnabel2017, Zhang2019a} and EP4s~\cite{Sayyad2022, Crippa2021} were discussed and few experiments exist revealing the existence of EP3s~\cite{Hodaei2017, Jing2017, Laha2020} and EP4s~\cite{Ding2016, Jin2018}, the generic features of EP3s and other higher-order EPs in 2D parameter space are not yet thoroughly investigated.

In this work, we show that in addition to the already abundant EP2s only EPs of order $n=3,4$ and $5$ can be generically induced by symmetries in 2D.
We find those symmetry-induced EPs appear in pairs in periodic parameter spaces.
Upon further analyzing the spectral structure of periodic $n$-band models, we find the following:
For $n=3$, we show that $\mathcal{PT}$, psH, and $\mathcal{CP}$ symmetry as well as CS have a very similar effect on the spectrum. In the presence of these symmetries, the EP3s, which scale as $\sim k^{1/3}$~\cite{Sayyad2022}, are intersected by a closed curve formed by EP2s. The EP2 curve forms the boundary of a three-level i-FS (FS) on the outside (inside) with $\mathcal{PT}$ and psH symmetry ($\mathcal{CP}$ symmetry and CS), whereas a two-level FS (i-FS) appears on the inside (outside), which is intersected by a three-level FA (i-FA) connecting the EP3s. The presence of SLS and psCS results in a drastically different phenomenology. In this case, the spectrum can be viewed as the two-band case with an additional flat band. Indeed, the EP3s scale as $\sim k^{1/2}$~\cite{Sayyad2022, Mandal2021}, and are connected via three-level (i)-FAs. We further show that whereas the EP3s can be demoted to EP2s in fine-tuned examples in the presence of SLS, EP2s find no room to arise in any two-band model with psCS.

In the case of $n=4$ and $5$, we find the emergence of EP4s and EP5s with rich spectral features in the presence of any two symmetries with different spectral constraints. For $n=4$, besides the EP4, we find second-order exceptional lines, four-level and two-level (i-)FSs as well as two-level (i-)FAs. The EP5 case amounts to the EP4 case with an additional flat band, and as such similar spectral features are found  with an increased degree of degeneracy.

Our results straightforwardly generalize to non-periodic 2D parameter spaces relevant for experiments. In this case,
the EPs do not have to appear in pairs anymore. Instead, in the experimentally accessible parameter space symmetry-protected EP3s, EP4s and EP5s might appear as single points with the same local spectral structure as for the EPs in the periodic systems.
All the EP lines and FAs are then promoted to open arcs, whereas the bounded FSs will appear as unbounded surfaces.

To explicitly show our results, we provide a minimal example for each type of structure. Our results characterize symmetry-induced EPs in 2D parameter space completely. As such, this work provides a significant contribution to the study of symmetry-protected NH phases and higher-order EPs, while they are also highly relevant for experiments, which are often conducted in 2D~\cite{Ozdemir2019}.

This paper is organized as follows. In Sec.~\ref{sec:gen_con}, we provide a general discussion on finding EPs in 2D. We then focus on EP3s in Sec.~\ref{sec:EP3s}, and EP4s and EP5s in Sec.~\ref{sec:EP4_EP5}. Section~\ref{sec:disc_con} presents a discussion and conclusion.

\begin{table}
    \centering
    \caption{Definitions of local (anti-)unitary symmetries}\label{tab:appendix_symmetry}
    \begin{tabular}{p{0.19\linewidth} p{0.45\linewidth} p{0.33\linewidth}}
        \hline \hline 
        Symmetry  & Symmetry constraint & energy constraint\\
        \hline
        $\mathcal{PT}$ & $H(\bm{k}) = \mathcal{A} H^*(\bm{k}) \mathcal{A}^{-1}$ & $\{\epsilon(\bm{k})\} = \{\epsilon^*(\bm{k})\}$ \\
        psH & $H(\bm{k}) = \varsigma H^\dagger(\bm{k}) \varsigma^{-1}$ & $\{\epsilon(\bm{k})\} = \{\epsilon^*(\bm{k})\}$ \\
        $\mathcal{CP}$ & $H(\bm{k}) = -\Theta H^*(\bm{k}) \Theta^{-1}$ & $\{\epsilon(\bm{k})\} = \{-\epsilon^*(\bm{k})\}$ \\
        CS & $H(\bm{k}) = -\Gamma H^\dagger(\bm{k}) \Gamma^{-1}$ & $\{\epsilon(\bm{k})\} = \{-\epsilon^*(\bm{k})\}$ \\
        psCS & $H^T(\bm{k}) = -X H(\bm{k}) X^{-1}$ & $\{\epsilon(\bm{k})\} = \{-\epsilon(\bm{k})\}$ \\
        SLS & $H(\bm{k}) = -\mathcal{S} H(\bm{k}) \mathcal{S}^{-1}$ & $\{\epsilon(\bm{k})\} = \{-\epsilon(\bm{k})\}$ \\
        \hline \hline
    \end{tabular}
    {\raggedright Here the unitary operator $U \in \{\varsigma,\Gamma,\mathcal{S},X\}$ satisfies $U^2=1$, while the unitary operator $A \in \{\mathcal{A}, \Theta\}$ obeys $AA^*=1$. \par}
\end{table}

\section{General Considerations} \label{sec:gen_con}

In order for an $n$-band system to exhibit an EP$n$ all terms except the leading one in the characteristic polynomial have to vanish.
If we set $\tr[H]=0$, which is simply a shift in the spectrum, we can express the characteristic polynomial in terms of the determinant and $n-2$ different traces~\cite{Curtright2012}: $\det[H]=0$ and $\tr[H^k]=0$ with $k=2,...,n-1$, which can be cast as $2(n-1)$ real constraints, which need to be simultaneously enforced in order to find an EP$n$~\cite{Sayyad2022}.
From now on we set $\tr[H]=0$ in all our models.
For brevity we refer to Ref.~\citenum{Sayyad2022} for the general case, while we here use specific characteristic polynomials for our $n$-band models.

The number of the real constraints can be reduced by imposing symmetries on the system.
From all unitary and anti-unitary symmetries only those acting local in momentum space reduce the number of constraints for finding an EP$n$.
It is shown in Ref.~\citenum{Sayyad2022} that these symmetries are $\mathcal{PT}$, psH and $\mathcal{CP}$ symmetry, as well as SLS, CS and psCS, which are defined in Table~\ref{tab:appendix_symmetry}. 
We note that $\mathcal{CP}$ symmetry is sometimes referred to as anti-$\mathcal{PT}$ symmetry in the literature~\cite{Bergman2021}.
Each symmetry is defined here in terms of some unitary generator.
For any choice of generator we find different allowed contributions to the Hamiltonian.
The interpretation of the symmetry then depends on the specific generator chosen, but we will not focus on this, but rather work out generic features of all symmetric Hamiltonians.
A specific choice of generator and Hamiltonian is only made to construct exemplary Hamiltonians, with which we show the spectral features, but we stress that we made an arbitrary choice of non-trivial unitary generator matrix there.
We emphasize that only the conditions imposed on the spectrum are important for the study of any type of degeneracy.
Thus the symmetries can be separated into three pairs, namely, $\mathcal{PT}$ and psH symmetry, $\mathcal{CP}$ symmetry and CS, and psCS and SLS.
The spectral constraint of each pair and the remaining constraints on the occurrence of EP$n$s in $n$-band systems are derived in Ref.~\citenum{Sayyad2022} and listed in Table~\ref{tab:symmetries}.

The mechanism of finding higher-order EPs in 2D is as follows: Two constraints can be generically fulfilled without any fine-tuning in a generic 2D parameter space.
Assuming periodicity, each constraint defines a closed curve, such that at the intersections of the curves EPs occur pairwise.
If we impose symmetries on the system that reduce the number of constraints for finding EP$n$s to two, EP$n$s thus generically appear in pairs in 2D.

Even though we focus on periodic parameter spaces in this work, our results can be straightforwardly generalized to non-periodic parameter spaces.
This can be seen from the fact that the reduction of the number of constraints due to symmetries works independent of the type of parameter space.
In non-periodic parameters spaces, the constraints define curves that are not necessarily closed.
If open curves intersect in the parameter space, we may find single symmetry-induced EPs.
If a constraint curve is closed instead we find EPs stabalized on these curves must appear in pairs.
In the following we will focus on the periodic two-dimensional parameter space. However, every spectral structure found there can be generalized to the non-periodic parameter space.
This is done by promoting all closed arcs and surfaces to open ones for single EPs.
In any experiment the local spectral structure at an EP would be probed, and this is identical for periodic and non-periodic parameter spaces.

In the presence of a single symmetry only EP3s can be generically realized in two dimensions, cf. Table~\ref{tab:symmetries}.
However, if two symmetries with different spectral constraints are simultaneously enforced on a system, the number of constraints one needs to impose to obtain an EP is further decreased, cf. the last row of Table~\ref{tab:symmetries}.
Thus EP4s and EP5s also occur in 2D in the presence of two different symmetries, whereas EP$n$s with $n\geq6$ cannot be generically induced by (anti-)unitary symmetries in 2D.

In this work, we use the strict definition that an EP$n$ is defined as having an $n$-fold spectral degeneracy accompanied by the coalescence of $n$ eigenvectors onto one, such that we define a threefold degeneracy at which two eigenvectors coalesce as an EP2, cf. Appendix C~\cite{supp}. 
This interpretation is further supported by the different geometric phases picked up when encircling an EP2 and an EP3~\cite{Demange2011}.

\begin{table}
    \centering
    \caption{Number of constraints for realizing EP$n$s in $n$-band systems restricted by local (anti-)unitary symmetries}\label{tab:symmetries}
    \begin{tabular}{p{0.257\linewidth}|p{0.342\linewidth}|p{0.351\linewidth}}
        \hline \hline 
        Symmetry  & \multicolumn{2}{l}{Number of constraints} \\ \cline{2-3}
        $[\textrm{spectrum}]$ & $n \in $ even & $n \in$ odd \\
        \hline \hline
        \begin{tabular}{@{}l@{}}$\mathcal{PT}$/psH \\ $[\{\epsilon\}=\{\epsilon^*\}]$ \end{tabular} & $n-1$~
        $\begin{cases}
            \Re[\det[{\cal H}]] , \\
            \Re[\tr[{\cal H}^k]] .
        \end{cases}$  & $n-1$~
       $\begin{cases}
            \Re[\det[{\cal H}]], \\
            \Re[\tr[{\cal H}^k]] .
        \end{cases}$ \\
        \hline
        \begin{tabular}{@{}l@{}}$\mathcal{CP}$/CS \\ $[\{\epsilon\}=\{-\epsilon^*\}]$ \end{tabular} & $n-1
        \begin{cases}
            \Re[\det[{\cal H}]] , \\
            \Re[\tr[{\cal H}^{l}]] ,\\
            \Im[\tr[{\cal H}^{m}]] .
        \end{cases}$  & $n-1$~
       $\begin{cases}
            \Im[\det[{\cal H}]] , \\
            \Re[\tr[{\cal H}^{l}]] , \\
            \Im[\tr[{\cal H}^{m}]] .
        \end{cases}$ \\
        \hline 
        \begin{tabular}{@{}l@{}}psCS/SLS \\ $[\{\epsilon\}=\{-\epsilon\}]$ \end{tabular} & $n
        \begin{cases}
            \det[{\cal H}] , \\
            \tr[{\cal H}^{l}] .
        \end{cases}$  & $n-1$~
       $\begin{cases}
            \tr[{\cal H}^{l}] .
        \end{cases}$ \\
        \hline 
        \begin{tabular}{@{}l@{}}combined$^1$ \\ $[\{\epsilon\}=\{\epsilon^*\} \, \wedge $ \\ $\, \{\epsilon\}=\{-\epsilon\}]$ \end{tabular} & $\frac{n}{2}
        \begin{cases}
            \Re[\det[{\cal H}]] ,\\
            \Re[\tr[{\cal H}^{l}]] .
        \end{cases}$  & $\frac{n-1}{2} 
        \begin{cases}
            \Re[\tr[{\cal H}^{l}]] .
        \end{cases}$ \\
        \hline \hline
    \end{tabular}
    {\raggedright Here $k\in \{1, \ldots n\}$, $l \in \{2 \leq l < n, l\in\textrm{even} \}$ and $m \in \{3 \leq m < n, m\in\textrm{odd} \}$. Behind the number of constraints we write the specific quantities that need to be set to zero to find EP$n$s. $^1$Here combined encompasses the constraints enforced by any pair of symmetries above, where the individual symmetries have different spectral constraints.\par}
\end{table}

\section{Exceptional points of order three} \label{sec:EP3s}

\begin{figure*}
    \centering
    \includegraphics[width=\textwidth]{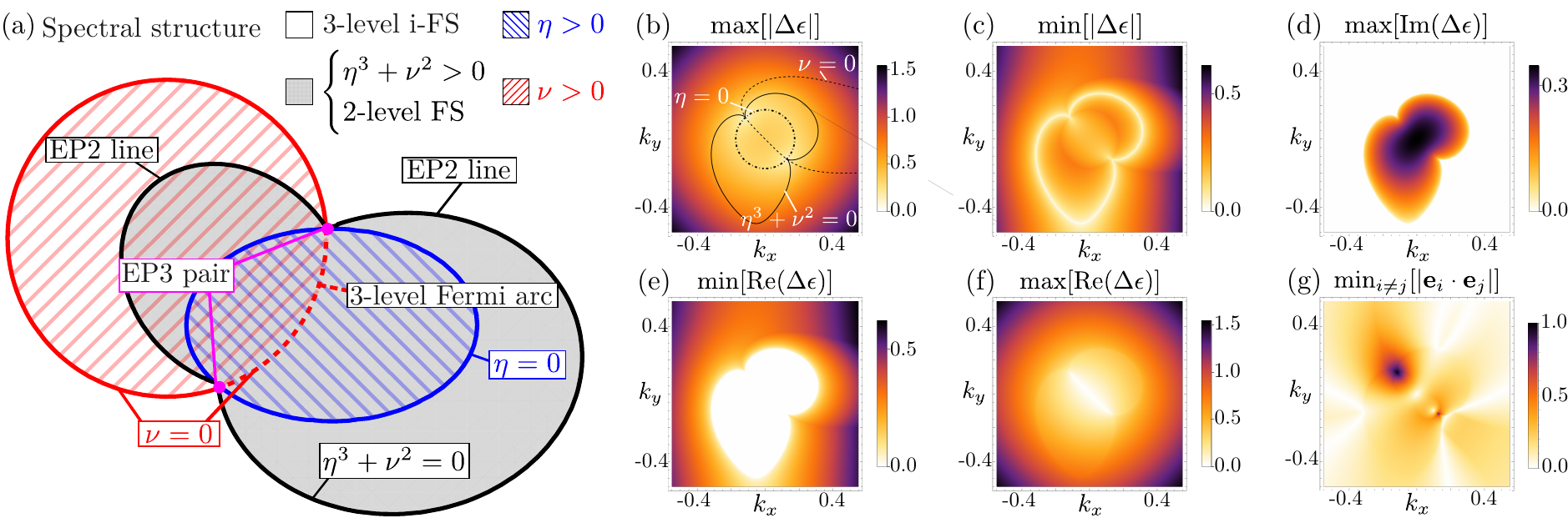}
    \caption{$\mathcal{PT}$/psH-symmetry-induced EP3s. (a) Sketch of the generic spectral structure around a $\mathcal{PT}$/psH-symmetry-induced pair of EP3s. Note that $\eta^3+\nu^2, \eta, \nu < 0$ outside the black, blue and red curves, respectively. $[$(b)-(g)$]$ Spectral structure of the $\mathcal{PT}$-symmetric model defined in Eq.~(\ref{eq:PT}) with $\xi=0.1$: (b) shows the maximum of the absolute value of the complex gap $\Delta \epsilon$, which disappears at the EP3 pair. On top, we plot the curves $\eta, \nu, \eta^3 + \nu^2 = 0$; (c) highlights the minimum of $|\Delta \epsilon|$, at which EP2 lines appear connecting the EP3 pair; in (d) and (e) the maximum of $\Im(\Delta \epsilon)$ and the minimum of $\Re(\Delta \epsilon)$ emphasize the three-level i-FS and the two-level FS, respectively; (f) highlights the three-level FA separating the two two-level FS; and (g) shows the minimum overlap of any pair of eigenvectors, which is one at the EP3s.}
    \label{fig:PT}
\end{figure*}

To study the behavior of EP3s in three-band systems we decompose the Hamiltonian in terms of the traceless, linearly independent Gell-Mann matrices $M^a,\textrm{ with }a=1,...,8$, which are generators of SU(3) with the property $M^a=(M^a)^\dagger$ (details are provided in Appendix A~\cite{supp}).
Any three-band Hamiltonian is given by
$H(\bm{k}) = \bm{h}(\bm{k}) \cdot \bm{M}$, where $\bm{M}=(M^1, M^2,...,M^8)^T$ is the vector of the Gell-Mann matrices and $\bm{h}(\bm{k})$ are complex-valued parameters that can be written as $\bm{h}(\bm{k})=\bm{h}_R(\bm{k})+i\bm{h}_I(\bm{k})$.
We introduce
\begin{align}
    \nu=\det[H]/2 \quad \textrm{and} \quad \eta= -\tr[H^2]/6,
\end{align}
such that characteristic polynomial can be expressed as $\mathcal{P}_3 = \epsilon^3 + 3\eta\epsilon - 2\nu$.

Degeneracies are obtained by setting the discriminant of $\mathcal{P}_3$ to zero, i.e., $\mathcal{D}_3=-108(\eta^3+\nu^2) = 0$.
An EP3 occurs iff $\nu=\eta=0$, while we note that EP2s appear when $\eta^3+\nu^2=0$ \cite{Patil2022}.
The closed curves defined by $\nu=0$, $\eta=0$ and $\eta^3+\nu^2=0$ divide the parameter space into different regions with different spectral structures.
This structure around the EP3s depends on the symmetry that induces the EPs, as we will see in the following.

\subsection{$\mathcal{PT}$ and psH-symmetry induced EP3s}

In the presence of $\mathcal{PT}$ or psH symmetry, either all eigenvalues are real, or one is real and the other two appear as complex conjugate pairs.
This results in $\nu,\eta\in\mathbb{R}$.
Using Cardano's method we can diagonalize the Hamiltonian.
With $\alpha_\pm = (\nu \pm \sqrt{\eta^3+\nu^2})^{1/3}$ and $\beta=(1+i\sqrt{3})/2=\exp(i\pi/3)$ the three eigenvalues are given by
\begin{equation}
    \begin{split}
        \epsilon_1 &= \alpha_+ + \alpha_-, \quad
        \epsilon_2 = -\beta^* \alpha_+ - \beta \alpha_-, \\
        \epsilon_3 &= -\beta \alpha_+ - \beta^* \alpha_-.
    \end{split}
\end{equation}

Let us start by considering the closed EP2 line $\eta^3+\nu^2=0$, which contains the pair of EP3s at $\nu=0=\eta$.
All other points on this line yield $\alpha_\pm = \sqrt[3]{\nu}\neq0$, such that $\epsilon_1\neq\epsilon_2=\epsilon_3$ with all eigenvalues real. 
The pair of EP3s are thus connected by two EP2 arcs, which form the boundaries to the regions $\eta^3+\nu^2>0$ and $<0$ on the inside and outside, respectively. 
If $\eta^3+\nu^2>0$ and $\nu \neq 0$, then $\alpha_\pm \in \mathbb{R}$, and the eigenvalues are $\epsilon_2=\epsilon_3^* \in \mathbb{C}$ and $\Re(\epsilon_{2/3}) \neq \epsilon_1 \in \mathbb{R}$. 
Since $\Re(\epsilon_2) = \Re(\epsilon_3)$ we obtain a two-level FS.
On the line $\nu = 0$, we find that $\eta^3+\nu^2>0$ implies $\eta>0$, and we obtain $\alpha_\pm=\pm\sqrt{\eta}$ such that $\epsilon_1=0$ and $\epsilon_2=i\sqrt{3\eta}=\epsilon_3^*\in i \mathbb{R}$. Therefore, the real part of all three eigenvalues coincide and this line corresponds to a three-level FA, separating the two-level FS and connecting the EP3s.
Considering $\eta^3+\nu^2<0$ instead, which implies $\eta<0$, we obtain $\alpha_+=\alpha_-^*$ independent of the sign of $\nu$.
As such, $\epsilon_i \in \mathbb{R}$ and $\epsilon_1\neq\epsilon_2\neq\epsilon_3$.
Since $\Im(\epsilon_i)=0$ for all three eigenvalues, the region $\eta^3+\nu^2<0$ forms a three-level i-FS. We show all these features in Fig.~\ref{fig:PT}(a).

To explicitly show the appearance of these generic features, let us introduce a $\mathcal{PT}$-symmetric model
\begin{align}
    \begin{split}
        H_\mathcal{PT}(\bm{k})  = &\sin(k_x) M^2 + h_s M^3 + \sin(k_y) M^4  \\ &+ i\xi \, \left(M^1+M^5+M^6\right)
    \end{split}\label{eq:PT} \\
        = &\begin{pmatrix}
        h_s & -i(\sin(k_x)-\xi) & \sin(k_y)+\xi \\
        i(\sin(k_x)+\xi) & -h_s & i\xi \\
        \sin(k_y)-\xi & i\xi&0
    \end{pmatrix} 
\end{align}
with $h_s=2-\cos(k_x)-\cos(k_y)$.
This model obeys $\mathcal{PT}$ symmetry $H_\mathcal{PT}(\bm{k})=\mathcal{A}H^*_\mathcal{PT}(\bm{k})\mathcal{A}^{-1}$ with the generator $\mathcal{A}=\frac{\mathbb{1}_3}{3} + M^3 - \frac{M^8}{\sqrt{3}}=\text{diag}(1,-1,1)$.
In Figs.~\ref{fig:PT}(b)-(g), we plot the EP3s, EP2 lines, FSs and FAs for this model, and we see all the predicted features.
In Appendix B~\cite{supp} we show the same features appear for a psH-symmetric model.

\subsection{$\mathcal{CP}$-symmetry and CS-induced EP3s}

Here we study three-band systems with $\mathcal{CP}$ symmetry or CS~\footnote{We note that even though strictly speaking there is no CS with $n \in \textrm{odd}$, we nevertheless consider this case here in line with the NH literature.}. In this case, the eigenvalues are either purely imaginary, or one is imaginary and the other two appear as pairs mirrored along the imaginary axis.
Therefore, the constraints are $\nu\in i\mathbb{R}$, $\eta\in\mathbb{R}$ and $\eta^3+\nu^2\in\mathbb{R}$. As $\nu$ is imaginary in this case, we have to choose a different branch in the third root as compared to the case with $\mathcal{PT}$ or psH symmetry. We introduce $\gamma_\pm = \pm (\pm\nu + \sqrt{\eta^3+\nu^2})^{1/3}$, such that the three eigenvalues read
\begin{equation}
    \begin{split}
        \epsilon_1 &= \gamma_+ + \gamma_-, \quad
        \epsilon_2 = -\beta^* \gamma_+ - \beta \gamma_-, \\
        \epsilon_3 &= -\beta \gamma_+ - \beta^* \gamma_-.
    \end{split}
\end{equation}

\begin{figure*}
    \centering
    \includegraphics[width=\textwidth]{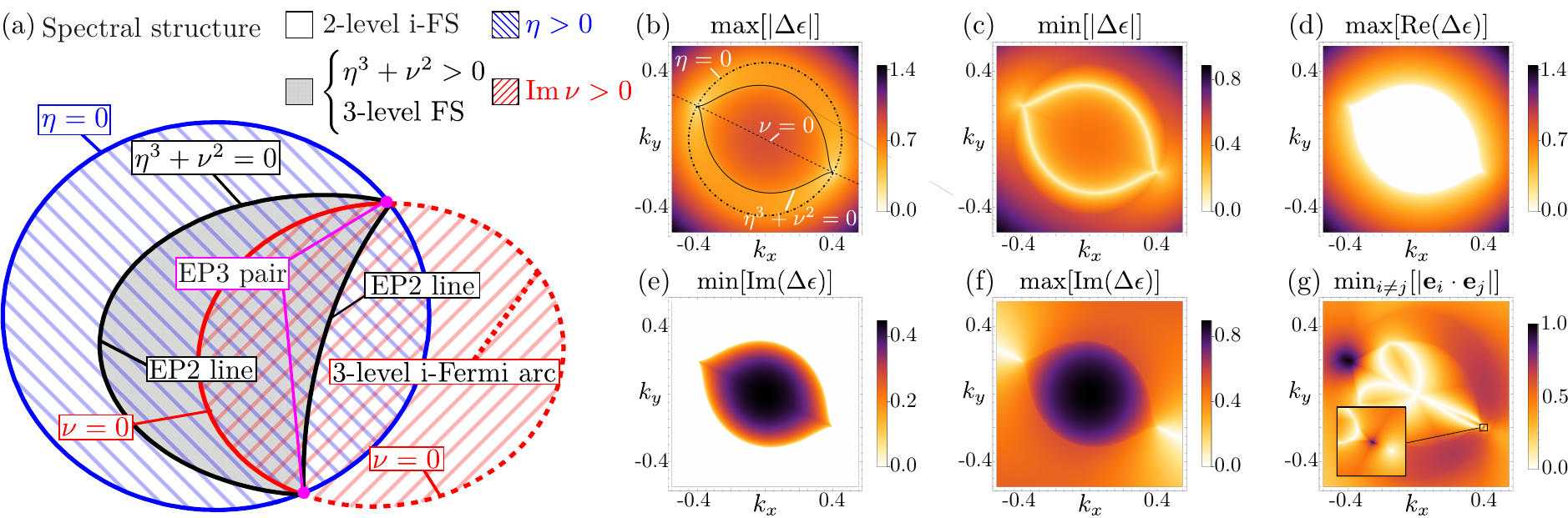}
    \caption{$\mathcal{CP}$-symmetry/CS-induced EP3s. (a) Sketch of the generic spectral structure around a $\mathcal{CP}$-symmetry/CS-induced pair of EP3s. Note that $\eta^3+\nu^2, \eta, \Im (\nu) < 0$ outside the black, blue and red curves, respectively. $[$(b)-(g)$]$ Spectral structure of the $\mathcal{CP}$-symmetric model in  Eq.~(\ref{eq:CP}) with $\xi=0.2$: (b) highlights the EP3 pair at which the maximum of $|\Delta \epsilon|$ disappears. Further the curves $\eta, \nu, \eta^3 + \nu^2 =0$ are plotted; (c) emphasizes the EP2 lines connecting the EP3 pair by showing the minimum of $|\Delta \epsilon|$; (d) and (e) show the three-level FS and the two-level i-FS, respectively; (f) highlights the three-level i-FA separating the two two-level i-FS; and (g) shows the minimum overlap of any pair of eigenvectors to verify that the threefold degeneracies are truly EP3s.}
    \label{fig:CP}
\end{figure*}

On the EP2 line $\eta^3+\nu^2=0$, we find $\gamma_+=-\beta \gamma_-$ for $\Im(\nu)>0$ and $\gamma_+=-\beta^* \gamma_-$ for $\Im(\nu)<0$.
Both cases correspond to two coalescing eigenvalues, while the third eigenvalue is different as long as $\nu \neq 0$.
Thus the pair of EP3s is connected by two arcs of EP2s as before. Now, the regions $\eta^3+\nu^2>0$ and $<0$ lie on the out- and inside of $\eta^3+\nu^2=0$, respectively.
For $\eta^3+\nu^2>0$ the fact that $\nu^2<0$ implies $\eta>|\nu|^{2/3}>0$.
Independent of the sign of $\Im(\nu)$ we obtain $\gamma_+=-\gamma_-^*$, which leads to three purely imaginary eigenvalues $\epsilon_i \in i\mathbb{R}$.
Thus we find a three-level FS in this region of the parameter space.
If $\eta^3+\nu^2<0$ and $\nu\neq0$, then $\gamma_\pm\in\mathcal{C}$, and there is no general relation between $\gamma_+$ and $\gamma_-$.
In this case, the spectrum obeys $\epsilon_2=-\epsilon_3^*$ and $\Im(\epsilon_2)\neq\epsilon_1\in i\mathbb{R}$.
Therefore, the eigenvalues form a two-level i-FS in this region.
For $\nu=0$ in $\eta^3+\nu^2<0$ we can write $\gamma_\pm =\pm\sqrt{|\eta|}\exp(i\pi/6)$ and thus obtain $\epsilon_1=-\epsilon_2\in\mathbb{R}$ and $\epsilon_3=0$.
Thus the imaginary parts of all eigenvalues coincide and this yields a three-level FA separating the two-level FS. The exceptional arcs and Fermi structure of a general $\mathcal{CP}$-symmetric or CS model is sketched in Fig.~\ref{fig:CP}(a).

To show the predicted features, we introduce a $\mathcal{CP}$-symmetric model Hamiltonian
\begin{align}
    \begin{split} 
        H_\mathcal{CP}(\bm{k})  = &\sin(k_x) M^1 + h_s M^5 + \sin(k_y) M^6 \\ 
        &+ i\xi \, \left(2M^2+M^7\right) 
    \end{split}\label{eq:CP} \\
    =&\begin{pmatrix}
        0 & \sin(k_x)+2\xi & -ih_s \\
        \sin(k_x)-2\xi & 0 & \sin(k_y)+\xi \\
        ih_s & \sin(k_y)-\xi & 0
    \end{pmatrix}
\end{align}
with $h_s=2-\cos(k_x)-\cos(k_y)$.
The $\mathcal{CP}$ symmetry is generated by $\Theta=\frac{\mathbb{1}_3}{3} + M^3 - \frac{M^8}{\sqrt{3}}=\text{diag}(1,-1,1)$, and enforced by the constraint $H_\mathcal{CP}(\bm{k})=-\Theta H^*_\mathcal{CP}(\bm{k})\Theta^{-1}$.
The generic spectral features of this model can be observed in Figs.~\ref{fig:CP}(b)-(g). 
Identical features emerge for a CS model, which we show in Appendix B~\cite{supp}.

$\mathcal{CP}$ symmetry and CS have a very similar effect on the band structure as $\mathcal{PT}$ and psH symmetry: 
Indeed, also in the presence of these symmetries, EP3s appear in pairs connected via EP2 lines. 
However, due to the additional minus sign in the constraints on the eigenvalues, cf. Table~\ref{tab:symmetries}, all the open Fermi structures as shown in Fig.~\ref{fig:PT}(a) are now i-Fermi structures and vice versa. 
Moreover, those structures that appeared on the in(out)side of the EP2 curve now appear on the out(in)side.

\subsection{psCS and SLS-induced EP3s}

Due to the spectral symmetry and the odd number of bands, one eigenvalue of a psCS or SLS three-band system is always $\epsilon=0$.
Therefore we obtain $\nu=\det[H]/2=0$, which simplifies the expression of the remaining eigenvalues.
The diagonalization of the Hamiltonian yields
\begin{equation}
    \begin{split}
        \epsilon_0 = 0, \;\;
        \epsilon_+ = + i \sqrt{3} \sqrt{\eta}, \;\;
        \epsilon_- = -i \sqrt{3} \sqrt{\eta}.
    \end{split}
\end{equation}
Since $\eta\in\mathbb{C}$, there are two real constraints that need to be imposed on the system.
These constraints follow from the generic constraint $\eta=0$, and can be written as $\bm{d}_R^2-\bm{d}_I^2=0$ and $\bm{d}_R\cdot\bm{d}_I=0$.
The curve given by $\bm{d}_R\cdot\bm{d}_I=0$ defines two rather simple FAs.
In the region with $\bm{d}_R^2-\bm{d}_I^2>0$, the eigenvalues are purely real and we obtain a three-level i-FA.
However, if $\bm{d}_R^2-\bm{d}_I^2<0$, the eigenvalues are purely imaginary and the curve defines a three-level FA.
In three-band systems with psCS and SLS there are no more generic spectral features in 2D.

These features are captured by the three-band example Hamiltonian with psCS that reads
\begin{align}
    \begin{split} 
    H_\textrm{psCS}(\bm{k})  = &\sin(k_x) M^1 + h_s M^5 + \sin(k_y) M^6 \\ 
    &+ i\xi \, \left(M^1+M^5+M^6\right)
    \end{split}\label{eq:psCS} \\
    =&\begin{pmatrix}
        0 & \sin(k_x) +i \xi & -ih_s+\xi \\
        \sin(k_x) + i\xi & 0 & \sin(k_y) + i\xi \\
        ih_s-\xi & \sin(k_y) + i\xi & 0
    \end{pmatrix}
\end{align}
with $h_s=2-\cos(k_x)-\cos(k_y)$.
Here, the psCS is defined by $H^T_\textrm{psCS}(\bm{k})=-X H_\textrm{psCS}(\bm{k})X^{-1}$ with the generator $X=\frac{\mathbb{1}_3}{3} + M^3 - \frac{M^8}{\sqrt{3}}=\text{diag}(1,-1,1)$.
The spectral structure of this model is shown in Fig.~\ref{fig:psCS}. In Appendix B~\cite{supp} we show a model with SLS displaying the same spectral features.

It is important to note that the constraints to find EP3s in three-band systems with psCS/SLS are nearly identical to those for finding EP2s in two-band systems without symmetries~\cite{Bergholtz2021}. As such, these EP3s not only come in pairs connected via FAs but also display a square-root energy scaling~\cite{Sayyad2022}.
We further note that in the presence of psCS no EP2s can occur in three-band systems. 
The idea behind the proof of this no-go theorem is that one can show that psCS in two-band systems prevents the emergence of EP2s. Indeed, in this case $H({\bf k})$ only has one non-zero component with one of the two-dimensional Pauli matrices, i.e., $H({\bf k}) \sim \sigma_i$, such that an EP2 can never occur. In Appendix C~\cite{supp} we show in detail how this extends to three-band systems.
There is no such no-go theorem for SLS, where EP2s accompanied by an orthogonal flat band can occur in fine-tuned examples.
In this case, the spectrum looks identical to that of the EP3 case and the spectral winding numbers take equal values.
To define the spectral winding, we generalize the vorticity $v$ introduced in Refs.~\onlinecite{Leykam2017,Shen2018} to three-band systems with psCS and SLS by taking the difference of the two dispersive bands, i.e., $\Delta \epsilon (\bm{k})= \epsilon_+(\bm{k})-\epsilon_-(\bm{k})$.
The vorticity is then defined by $v=-\oint_\mathcal{C} \frac{d\bm{k}}{2\pi}\cdot\nabla_{\bm{k}} \textrm{arg}\left[\Delta \epsilon(\bm{k})\right]$ for a closed curve $\mathcal{C}$ that encircles a single EP, and can take the values $\pm1/2$.
The sum over all EPs in the system must be 0, and thus the charge of the EP3s of a single pair is opposite. We note that the EP3s with a flat band thus have the same vorticity as EP2s.
Moreover, the constraints for finding EP2s in a three-band model with SLS are identical to the constraints for finding EP3s. 
As such, one would have to calculate the Jordan decomposition to determine the order of the EPs. Alternatively, one could study the geometric phases picked up by the eigenvectors corresponding to the dispersive bands upon encircling an EP, which is different for an EP2 and an EP3 with a flat band~\cite{Demange2011}: Whereas one needs to encircle the EP twice to return to the initial eigenvector in both cases, a geometric phase of $\pi$ is picked up if the EP is of order 2, whereas no geometric phase is acquired in the EP3 case~\cite{Demange2011}.
In Appendix C~\cite{supp} we show a model with SLS, which hosts an EP2 pair.

\begin{figure}[t]
    \centering
    \includegraphics[width=\columnwidth]{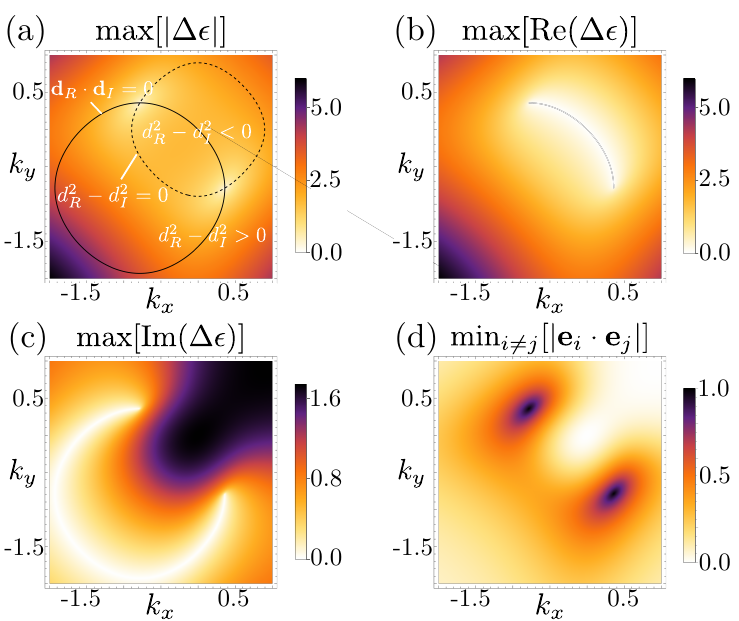}
    \caption{Spectral structure of psCS-induced EP3s in the model defined in Eq.~(\ref{eq:psCS}) with $\xi=0.5$. (a) highlights the EP3 pair, while also depicting the closed curves on which the constraints are fulfilled; (b) and (c) show the real and imaginary three-level FAs, respectively; and (d) displays the minimum overlap of eigenvector pairs of the model to verify the third-order EPs.}
    \label{fig:psCS}
\end{figure}

\section{Symmetry-induced Exceptional points of order four and five} \label{sec:EP4_EP5}

Lastly, we turn to 4- and five-band models in the presence of two symmetries with different spectral constraints realizing EP4s and EP5s, respectively. We will see that the general considerations and subsequent results are very similar for both these cases.

\subsection{Exceptional points of order four}

For four-band models the Hamiltonian can be decomposed in terms of generalized Gell-Mann matrices. These matrices $\Lambda^a$ with $a=1,...,15$ are the generators of SU(4), which are traceless, linearly independent matrices and fulfill $\Lambda^a=(\Lambda^a)^\dagger$ (see Appendix A~\cite{supp}).
A general four-band Hamiltonian can be written as $H(\bm{k}) = \bm{h}(\bm{k}) \cdot \bm{\Lambda}$, where $\bm{\Lambda}=(\Lambda^1, \Lambda^2,...,\Lambda^{15})^T$ is the vector of the generalized Gell-Mann matrices and $\bm{h}(\bm{k})$ is a complex-valued parameter vector and can be written as $\bm{h}(\bm{k})=\bm{h}_R(\bm{k})+i\bm{h}_I(\bm{k})$.
We introduce $\Tilde{\nu}=\det[H]$ and $\Tilde{\eta}=\tr[H^2]/4$ with $\Tilde{\nu},\Tilde{\eta}\in\mathbb{R}$, where we made use of the constraints on the eigenvalue in Table~\ref{tab:symmetries}.
The characteristic polynomial reads $\mathcal{P}_4 = \epsilon^4 - 2\Tilde{\eta}\epsilon^2 + \Tilde{\nu}$ with the discriminant given by $\mathcal{D}_4=64(\Tilde{\eta}^2-\Tilde{\nu})\Tilde{\nu}$, such that the eigenvalues read 
\begin{equation}
    \epsilon_{\pm_1,\pm_2} = \pm_1 \sqrt{\Tilde{\eta} \pm_2 \sqrt{\Tilde{\eta}^2-\Tilde{\nu}}}
\end{equation}
An EP4 is found when $\Tilde{\nu}=\Tilde{\eta}=0$. 
Since $\Tilde{\nu}=0$ and $\Tilde{\eta}=0$ are closed curves in 2D parameter space the EP4s indeed appear in pairs. 
The symmetry induced EP4s scale as $\sim k^{1/2}$ contrary to generic EP4s, which scale as $\sim k^{1/4}$.

\begin{figure*}
    \centering
    \includegraphics[width=\textwidth]{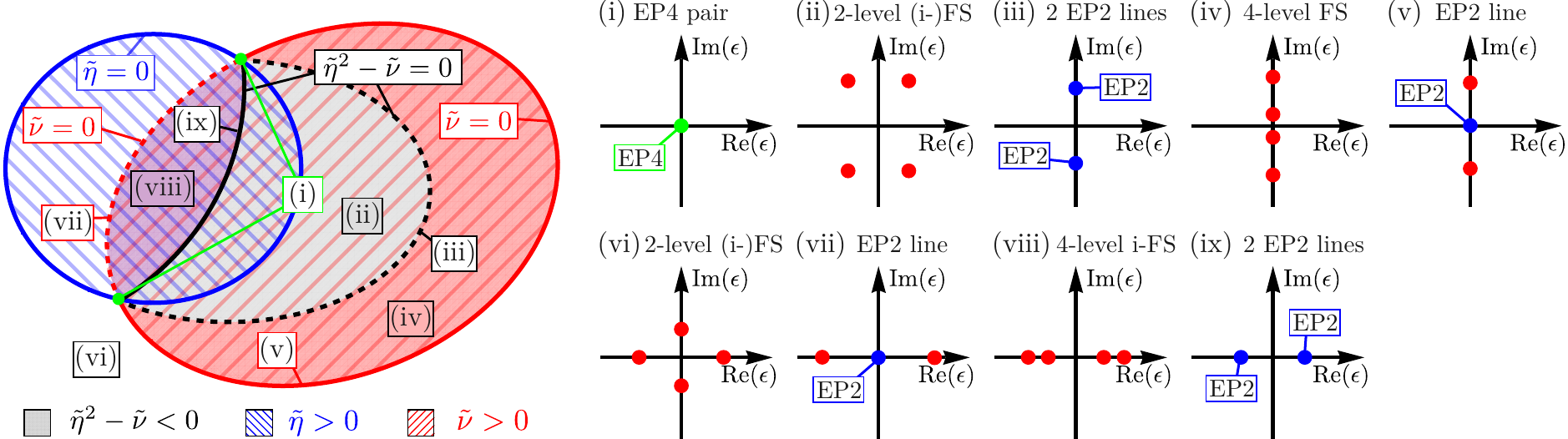}
    \caption{Spectral structure of symmetry-induced EP4s. On the left side there is a sketch of the different regions in 2D parameters space, which are defined by the constraints we have to impose to realize EPs. Each line and region is labeled with (i)-(ix), and on the right their generic complex spectrum is displayed. We note that (v) and (vii) also reveal a two-level FA and i-FA, respectively. The spectral structure of EP5s is similar to the structure of EP4s. By substituting $\Tilde{\nu}$ with $\kappa$ and adding a flat band at zero, we obtain the eigenvalue structure around an EP5 pair. The flat band promotes the EP4s to EP5s [(i)], the EP2 lines to EP3 lines [(v) and (vii)], the four-level (i-)FSs to five-level ones [(iv) and (viii)], and the two-level (i)-FAs and (i-)FSs to three-level ones [(v)-(vii)].}
    \label{fig:EP4}
\end{figure*}

Degeneracies appear in the spectrum if the discriminant $\mathcal{D}_4$ vanishes.
The arcs defined by $\mathcal{D}_4=0$ begin and end at the EP4s.

For $\Tilde{\nu}=0$ (red line in Fig.~\ref{fig:EP4}) and $\Tilde{\eta}\neq0$ (blue dashed region) there are always two eigenvalues $\epsilon_{\pm_1,-\textrm{sgn}(\Tilde{\eta})}=0$, which results in a two-level exceptional arc.
The other two eigenvalues are given by $\epsilon_{\pm_1,\textrm{sgn}(\Tilde{\eta})}=\pm_1\sqrt{2\Tilde{\eta}}$ and are therefore either purely real or imaginary, depending on the sign of $\Tilde{\eta}$: For $\Tilde{\eta}>0$ both eigenvalues are real and thus form a two-level i-Fermi arc, where the imaginary part coincides with the imaginary part of the exceptional arc, cf. Fig.~\ref{fig:EP4}(vii).
If $\Tilde{\eta}<0$ the eigenvalues are purely imaginary and the arc is a two-level Fermi arc with the same real energy as the exceptional arc, cf. Fig.~\ref{fig:EP4}(v).

On the arcs defined by $\Tilde{\eta}^2-\Tilde{\nu}=0$ (black line in Fig.~\ref{fig:EP4}) with $\Tilde{\nu}\neq0\neq\Tilde{\eta}$ the eigenvalue structure is different.
The eigenvalues are given by $\epsilon_{\pm_1,\pm_2}=\pm_1\sqrt{\Tilde{\eta}}$ and both eigenvalues are twofold degenerate. As $\Tilde{\eta}^2-\Tilde{\nu}=0$ amounts to satisfying one real constraint, we postulate that this arc corresponds to a two-level exceptional arcs.
Again the sign of $\Tilde{\eta}$ determines whether the eigenvalues are real or imaginary, cf. Figs.~\ref{fig:EP4}(ix) and (iii), respectively.

Considering $\Tilde{\eta}^2-\Tilde{\nu}>0$ (outside the black curve in Fig.~\ref{fig:EP4}) and  $\Tilde{\nu}>0$ (red dashed region), we find $|\Tilde{\eta}|>\sqrt{\Tilde{\eta}^2-\Tilde{\nu}}$, such that depending on the sign of $\Tilde{\eta}$ all four eigenvalues are either real or purely imaginary.
For $\Tilde{\eta}>0$ (blue dashed region) we thus obtain a four-level i-FS and for $\Tilde{\eta}<0$ (outside the blue curve) we obtain a four-level FS, cf. Figs.~\ref{fig:EP4} (viii) and (iv), respectively.
If $\Tilde{\nu}<0$ (outside the red curve) then $|\Tilde{\eta}| < \sqrt{\Tilde{\eta}^2-\Tilde{\nu}}$, such that two eigenvalues are real and two are purely imaginary.
Thus in this region there is always a two-level FS and a two-level i-FS, cf. Fig.~\ref{fig:EP4} (vi).

The condition $\Tilde{\eta}^2-\Tilde{\nu}<0$ (gray region in Fig.~\ref{fig:EP4}), on the other hand, implies $\Tilde{\nu}>0$.
All eigenvalues are truly complex, but due to the various spectral symmetry constraints there is a special eigenvalue structure.
We always obtain two FSs and two i-FSs with the real and imaginary part symmetric around zero, cf. Fig.~\ref{fig:EP4}(ii).
Each eigenvalue is part of one Fermi and one i-FS, but it shares each surface with a different other eigenvalue.

These predicted features can be realized with a four-band model on which $\mathcal{PT}$ symmetry and SLS are imposed.
The Hamiltonian is given by
\begin{align}
    \begin{split} 
    H_4(\bm{k})  = &\sin(k_x) \Lambda^1 + \frac{1}{2}\sin(k_y) \Lambda^7 + h_s \Lambda^{13}
    + i\xi \, \Lambda^9
    \end{split}\\
    =&\begin{pmatrix}
        0 & \sin(k_x) & 0 & i\xi \\
        \sin(k_x) & 0 & -i\sin(k_y)/2 & 0 \\
        0&  i\sin(k_y)/2 & 0 & h_s \\
        i\xi & 0 & h_s & 0
    \end{pmatrix}
\end{align}
with $h_s=2-\cos(k_x)-\cos(k_y)$ and $\xi=0.2$.
The generator for $\mathcal{PT}$ symmetry $H_4(\bm{k})=\mathcal{A}_4 H^*_4(\bm{k})\mathcal{A}_4^{-1}$ is $\mathcal{A}_4=\frac{2}{\sqrt{3}}\Lambda^8+\frac{\sqrt{2}}{\sqrt{3}}\Lambda^{15}=\text{diag}(1,1,-1,-1)$ and SLS is defined by $H_4(\bm{k})=-\mathcal{S}_4 H_4(\bm{k})\mathcal{S}_4^{-1}$ with the generator $\mathcal{S}_4=\Lambda^3-\frac{1}{\sqrt{3}}\Lambda^8+\frac{\sqrt{2}}{\sqrt{3}}\Lambda^{15}=\text{diag}(1,-1,1,-1)$.
In this model two pairs of EP4s occur and they are connected by exceptional arcs.
This is simply a generalization of the structure for a single pair.
The parameter space is divided into more regions due to multiple intersections of the curves defined by the constraints.
However, for each region the previous analysis holds and the spectral structure surrounding a single EP4 is as described before.

\subsection{Exceptional points of order five}

For the five-band case, we find the spectral structure nearly identical to the four-band case plus a zero-eigenvalue.
We again decompose the Hamiltonian in terms of the generalized Gell-Mann matrices that are generators of SU(5).
In terms of these traceless linearly independent matrices $W^a$ with $a=1,...,24$ a general four-band Hamiltonian can be written as $H(\bm{k}) = \bm{h}(\bm{k}) \cdot \bm{W}$, where $\bm{W}=(W^1, W^2,...,W^{24})^T$ is the vector of the Gell-Mann matrices and $\bm{h}(\bm{k})$ are complex-valued parameters that can be written as $\bm{h}(\bm{k})=\bm{h}_R(\bm{k})+i\bm{h}_I(\bm{k})$.
We define $\Tilde{\eta}=\tr[H^2]/4$ and $\kappa=\{[\tr(H^2)]^2-2\tr(H^4)\}/8$, such that the constraints on the eigenvalues lead to $\Tilde{\eta}, \kappa \in \mathbb{R}$.
The characteristic polynomial simplifies to a polynomial with only odd powers given by $\mathcal{P}_5 = \epsilon^5 - 2\Tilde{\eta}\epsilon^3 + \kappa\epsilon$.
The discriminant is given by $\mathcal{D}_5=64(\Tilde{\eta}^2-\kappa)\kappa$, and the eigenvalues read 
\begin{equation}
    \epsilon_0 = 0,
    \epsilon_{\pm_1,\pm_2} = \pm_1 \sqrt{\Tilde{\eta} \pm_2 \sqrt{\Tilde{\eta}^2-\kappa}}\, .
\end{equation}
Since the characteristic polynomial is a biquadratic fourth-order polynomial multiplied with $\epsilon$ we obtain formally the solutions of the four-band case with the addition of a flat zero-energy level.

To study the generic features of a five-band model subject to two symmetries with different constraints, we note that the same logic applies here as for the four-band case discussed above because of the similarity of the eigenvalue equations.
Indeed, the role played by $\tilde{\nu}$ for the four-band case is now played by $\kappa$. For $\kappa=0$, the second-order exceptional lines at zero energy found on the curve spanned by $\tilde{\nu} = 0$ in the four-band model, cf. Figs.~\ref{fig:EP4}(v) and (vii), are promoted to order three.
Both four-level (i-)FS, cf. Figs.~\ref{fig:EP4}(iv) and (viii), are promoted to five-level surfaces, and the two two-level FS for $\Tilde{\eta}<0$ and $\Tilde{\nu}<0$, cf. Fig.~\ref{fig:EP4}(vi), are three-level surfaces in the five-band case for $\kappa<0$, where the flat band is part of both.
Otherwise the spectral features are not affected by the addition of the flat band, and the features pointed out in Figs.~\ref{fig:EP4}(ii), (iii), and (ix) remain unchanged.

The predicted features can be realized with a five-band model on which $\mathcal{PT}$ symmetry and SLS are imposed.
The Hamiltonian is given by
\begin{widetext}
    \begin{align}
        H_5(\bm{k})  = &\sin(k_x) \left(W^2+W^{23}\right) + \sin(k_y) W^7 + h_s \left(W^{10}+W^{14}\right) + i\xi \, \left(W^1+W^{22}\right) \\
        =& \begin{pmatrix}
            0 & -i(\sin(k_x - \xi) & 0 & -ih_s & 0 \\
            i(\sin(k_x + \xi) & 0 & -\sin(k_y) & 0 & 0 \\
            0 & i\sin(k_y) & 0 & -ih_s & 0 \\
            ih_s & 0 & ih_s & 0 & i(\sin(k_x)-\xi) \\
            0 & 0 & 0 & i(\sin(k_x + \xi) & 0
        \end{pmatrix}
    \end{align}
\end{widetext}
with $h_s=2-\cos(k_x)-\cos(k_y)$ and $\xi=0.2$.
The generators for $\mathcal{PT}$ symmetry $H_5(\bm{k})=\mathcal{A}_5 H^*_5(\bm{k})\mathcal{A}_5^{-1}$ and SLS $H_5(\bm{k})=-\mathcal{S}_5 H_5(\bm{k})\mathcal{S}_5^{-1}$ are identical $\mathcal{A}_5=\mathcal{S}_5=\frac{\mathbb{1}_5}{5}+W^3-\frac{1}{\sqrt{3}}W^8+\frac{\sqrt{2}}{\sqrt{3}}W^{15}-\frac{\sqrt{2}}{\sqrt{5}}W^{24}=\text{diag}(1,-1,1,-1,1)$.
Here two pairs of EP5s occur, which are connected by exceptional arcs.
Again this is simply a generalization of the structure for a single pair, where the exceptional arcs terminate at different EPs.

Similar to our previous considerations of EP2s in three-band systems with SLS, we emphasize that it is not possible to distinguish EP5s from EP4s with an orthogonal flat band using the spectral structure alone.
In fact to any four-band model with symmetry-induced EP4s a flat band can be added without affecting the symmetry constraints. To subsequently determine whether the EP4 is rendered an EP5 or is still an EP4, one would again have to compute the Jordan decomposition.
Here we cannot construct the geometric phase of the eigenvectors upon encircling an EP4 or EP5, because of the EP lines originating at the respective EPs.

\section{Discussion and Conclusion} \label{sec:disc_con}

In this work, we exhaustively discussed the appearance of symmetry-induced higher-order EPs in 2D parameter space.
We showed that while several (anti-)unitary symmetries exist that reduce the codimension of the EPs, these symmetries can be divided into three groups based on their resulting spectral constraints. As such, the cases discussed in this work completely characterize all the possible symmetry-protected multi-band spectral features in 2D.
For EP3s we derived the full spectral structure depending on the underlying symmetry. We gave for each symmetry minimal examples featuring the induced higher-order EPs.
We note that for these examples we chose a convenient generator, and subsequently derived the Hamiltonian. We would like to emphasize that the choice of generator is indeed a choice. Instead, we could have chosen a different generator in which case we would have found a different example Hamiltonian obeying the relevant symmetry.

We showed that each pair of EPs is accompanied by a generic spectral structure, which generally includes exceptional arcs as well as open Fermi structures of various degrees.
Further we showed that due to multiple symmetries EP4s and EP5s are induced in 2D parameter space.
These EPs also have to appear in pairs, and the spectral structure around them is independent of the specific symmetries of the system.
We saw that symmetry-induced higher-order EPs often appear with characteristics very similar to lower-order EPs. 
Indeed, we find EP3s as well as EP4s and EP5s, which look like EP2s in the sense that they have a square-root dispersion. 
Nevertheless the spectral structure is a distinct feature that identifies the EPs according to the symmetries imposed on the system.
If a symmetry-induced EP is parametrically encircled, the eigenvalues exchange with each other in a manner fully determined by the symmetry.
We note that in the presence of $\mathcal{PT}$, psH, and $\mathcal{CP}$ symmetry as well as CS the presence of EP2 arcs connecting the EP3s hinders a proper definition of the topological charge of an EP3. Similarly, in the 4- and five-band cases, the EP4s and EP5s are connected via EP2 and EP3 lines, respectively, such that it is also not possible to define a charge. Therefore, the only instance in which one can compute a topological charge is in the three-band case with psCS or SLS.

The EPs discussed here can be probed in experiments described by a Hamiltonian $H$ with two tunable parameters that span the parameter space.
Upon probing an induced EP experimentally, the rich spectral structure predicted in this work can be utilised.
From the Hamiltonian $H$ describing the experiment the EP3s,  EP4s, and EP5s can be determined by requiring $\nu=\eta=0$, $\Tilde{\nu}=\Tilde{\eta}=0$, and $\kappa=\Tilde{\eta}=0$, respectively.
This tuning can be achieved by varying the two tuning parameters.
However, due to noise in the experimental setup and thus in the parameters of $H$ the EPs themselves might not be measurable directly.
Instead a single EP can be encirled parametrically by measuring the eigenvalue spectrum in parameter space on a closed loop surrounding the EP.
Depending on the symmetry in the system, a specific spectral pattern should be visible upon performing the loop. For example, in the case of a $\mathcal{PT}$- or psH-symmetric three-band Hamiltonian, this loop has to cross each two second-order EPs and a three-level Fermi arc originating at the EP3 once.
If these signatures are visible in the spectrum when performing the loop, there has to be at minimum a single symmetry-induced higher-order EP inside the parameter loop.
By contracting the loop the exact position of the EP might be measured.
We emphasize that the qualitative spectral structure does not depend on the length or the shape of the loop but only on the presence of a single EP, which is surrounded.

We note that three-band models with $\mathcal{PT}$ symmetry and SLS were also considered in Ref.~\citenum{Mandal2021}~\footnote{We note that in Ref.~\citenum{Mandal2021}, SLS is referred to as chiral symmetry, whereas what we call CS is not discussed in that work. Here, we use the definitions as in Refs.~\onlinecite{Kawabata2019, Sayyad2022}.}. 
There the authors identify the same features as in Fig.~\ref{fig:PT}(a) except for the three-level FA and i-FS, whereas for the SLS case their findings correspond to what we show in Fig.~\ref{fig:psCS}. 
Moreover, our work adds a nuance to the statement in Ref.~\citenum{Mandal2021}, where it is mentioned that no EP2 may occur in a three-band model with SLS. Here we show that while EP2s indeed do not generically appear in the presence of this symmetry, fine-tuned models can be found in which EP2s do arise as shown in an explicit example in Appendix C~\cite{supp}. 
Instead, we proof that there is no room for EP2s to arise in the presence of psCS, cf. Appendix C~\cite{supp}.

From Table~\ref{tab:symmetries} we see that EP3s can also be stabilized in 1D in the presence of two symmetries with different constraints. Similar to the EP5 case, this amounts to a situation very similar to having symmetry-induced EP2s in 1D with an additional flat band at zero.

This work not only provides full theoretical insight into symmetry-protected multi-band features in 2D but is also highly relevant for experiment, where non-Hermiticity finds many applications in dissipative metamaterials~\cite{Ozdemir2019,Opala2023}. Indeed, all the features discussed in this work are the only ones that could generically appear in 2D besides EP2s~\cite{Heiss2012, Zhen2015}, and we expect they can be straightforwardly engineered in a plethora of different experimental platforms, ranging from optical metasurfaces~\cite{Zhen2015} to optical fibres~\cite{Bergman2021} and microcavities~\cite{Peng2014}, where symmetry-protected EP3s have already been observed~\cite{Hodaei2017, Laha2020}.

\acknowledgments

We are grateful to Jacob Fauman and Emil J. Bergholtz for insightful discussions. We acknowledge funding from the Max Planck Society Lise Meitner Excellence Program 2.0.
We also acknowledge funding from the European Union via the ERC Starting Grant “NTopQuant”. 
Views and opinions expressed are however those of the authors only and do not necessarily reflect those of the European Union or the European Research Council (ERC). Neither the European Union nor the granting authority can be held responsible for them.

\bibliography{references.bib}

\appendix

\section{Generalized Gell-Mann matrices}\label{app:B}

The basis matrices of $n$-band systems are the generalized Gell-Mann matrices $\lambda^a$ with $a\in  \{1,...,n^2-1\}$ that span the Lie algebra of the SU($n$) group.
We note that the term Gell-Mann matrices is commonly used for the matrices associated with SU(3), while we use the generalization to higher order to describe four-band and five-band systems.
The matrices are traceless, i.e., $\tr[\lambda^a] = 0$, and self-adjoint $(\lambda^a)^\dagger=\lambda^a$.
They satisfy the relations
\begin{align}
    \left[\lambda^i,\lambda^j\right] &= 2if_{ijk}\lambda^k ,\\
    \left\{\lambda^i,\lambda^j\right\} &= \frac{4}{N}\delta_{ij} \mathbb{1}_n + 2d_{ijk}\lambda^k,
\end{align}
where $\mathbb{1}_n$ denotes the $n\times n$ identity matrix, and $d_{ijk}$ and $f_{ijk}$ the symmetric and anti-symmetric structure constants, respectively, defined by
\begin{align}
    d_{ijk} = \frac{1}{4}\tr\left(\lambda^i\{\lambda^j,\lambda^k\}\right) ,\\
    f_{ijk} = - \frac{i}{4}\tr\left(\lambda^i[\lambda^j,\lambda^k]\right).
\end{align}
In the main text we use the generalized Gell-Mann matrices for $n=3,4,5$.
The Gell-Mann matrices associated with SU(3) are 
\begin{align}
    M^1&=\begin{pmatrix}
        0&1&0\\
        1&0&0\\
        0&0&0
    \end{pmatrix} \; ,\;\; M^2=\begin{pmatrix}
        0&-i&0\\
        i&0&0\\
        0&0&0
    \end{pmatrix} \; ,\\
    M^3&=\begin{pmatrix}
        1&0&0\\
        0&-1&0\\
        0&0&0
    \end{pmatrix} \; ,\;\; M^4=\begin{pmatrix}
        0&0&1\\
        0&0&0\\
        1&0&0
    \end{pmatrix} \; ,\\
    M^5&=\begin{pmatrix}
        0&0&-i\\
        0&0&0\\
        i&0&0
    \end{pmatrix} \; ,\;\; M^6=\begin{pmatrix}
        0&0&0\\
        0&0&1\\
        0&1&0
    \end{pmatrix} \; ,\\
    M^7&=\begin{pmatrix}
        0&0&0\\
        0&0&-i\\
        0&i&0
    \end{pmatrix} \; ,\;\; M^8=\begin{pmatrix}
        \frac{1}{\sqrt{3}}&0&0\\
        0&\frac{1}{\sqrt{3}}&0\\
        0&0&\frac{-2}{\sqrt{3}}
    \end{pmatrix} \; .
\end{align}
The generalized Gell-Mann matrices spanning the SU(4) Lie algebra are given by
\begin{align}
    \Lambda^1&=\begin{pmatrix}
        0&1&0&0\\
        1&0&0&0\\
        0&0&0&0\\
        0&0&0&0
    \end{pmatrix} \; ,\;\; \Lambda^2=\begin{pmatrix}
        0&-i&0&0\\
        i&0&0&0\\
        0&0&0&0\\
        0&0&0&0
    \end{pmatrix} \; ,\\
    \Lambda^3&=\begin{pmatrix}
        1&0&0&0\\
        0&-1&0&0\\
        0&0&0&0\\
        0&0&0&0
    \end{pmatrix} \; ,\;\; \Lambda^4=\begin{pmatrix}
        0&0&1&0\\
        0&0&0&0\\
        1&0&0&0\\
        0&0&0&0
    \end{pmatrix} \; ,\\
    \Lambda^5&=\begin{pmatrix}
        0&0&-i&0\\
        0&0&0&0\\
        i&0&0&0\\
        0&0&0&0
    \end{pmatrix} \; ,\;\; \Lambda^6=\begin{pmatrix}
        0&0&0&0\\
        0&0&1&0\\
        0&1&0&0\\
        0&0&0&0
    \end{pmatrix} \; ,\\
    \Lambda^7&=\begin{pmatrix}
        0&0&0&0\\
        0&0&-i&0\\
        0&i&0&0\\
        0&0&0&0
    \end{pmatrix} \; ,\;\; \Lambda^8=\begin{pmatrix}
        \frac{1}{\sqrt{3}}&0&0&0\\
        0&\frac{1}{\sqrt{3}}&0&0\\
        0&0&\frac{-2}{\sqrt{3}}&0\\
        0&0&0&0
    \end{pmatrix} \; ,\\
    \Lambda^9&=\begin{pmatrix}
        0&0&0&1\\
        0&0&0&0\\
        0&0&0&0\\
        1&0&0&0
    \end{pmatrix} \; ,\;\; \Lambda^{10}=\begin{pmatrix}
        0&0&0&-i\\
        0&0&0&0\\
        0&0&0&0\\
        i&0&0&0
    \end{pmatrix} \; ,\\
    & \notag \\
    & \notag \\
    \Lambda^{11}&=\begin{pmatrix}
        0&0&0&0\\
        0&0&0&1\\
        0&0&0&0\\
        0&1&0&0
    \end{pmatrix} \; ,\;\; \Lambda^{12}=\begin{pmatrix}
        0&0&0&0\\
        0&0&0&-i\\
        0&0&0&0\\
        0&i&0&0
    \end{pmatrix} \; ,\\
    \Lambda^{13}&=\begin{pmatrix}
        0&0&0&0\\
        0&0&0&0\\
        0&0&0&1\\
        0&0&1&0
    \end{pmatrix} \; ,\;\; \Lambda^{14}=\begin{pmatrix}
        0&0&0&0\\
        0&0&0&0\\
        0&0&0&-i\\
        0&0&i&0
    \end{pmatrix} \; ,\\
    \Lambda^{15}&=\begin{pmatrix}
        \frac{1}{\sqrt{6}}&0&0&0\\
        0&\frac{1}{\sqrt{6}}&0&0\\
        0&0&\frac{1}{\sqrt{6}}&0\\
        0&0&0&\frac{-3}{\sqrt{6}}
    \end{pmatrix} \; .
\end{align}
\newpage
The generalized Gell-Mann matrices spanning the SU(5) Lie algebra are given by
\begin{widetext}
\begin{align}
    W^1&=\begin{pmatrix}
        0&1&0&0&0\\
        1&0&0&0&0\\
        0&0&0&0&0\\
        0&0&0&0&0\\
        0&0&0&0&0
    \end{pmatrix} \; , \quad
    W^2=\begin{pmatrix}
        0&-i&0&0&0\\
        i&0&0&0&0\\
        0&0&0&0&0\\
        0&0&0&0&0\\
        0&0&0&0&0
    \end{pmatrix} \; ,\quad
    W^3=\begin{pmatrix}
        1&0&0&0&0\\
        0&-1&0&0&0\\
        0&0&0&0&0\\
        0&0&0&0&0\\
        0&0&0&0&0
    \end{pmatrix} \; ,\\
    W^4&=\begin{pmatrix}
        0&0&1&0&0\\
        0&0&0&0&0\\
        1&0&0&0&0\\
        0&0&0&0&0\\
        0&0&0&0&0
    \end{pmatrix} \; ,\quad
    W^5=\begin{pmatrix}
        0&0&-i&0&0\\
        0&0&0&0&0\\
        i&0&0&0&0\\
        0&0&0&0&0\\
        0&0&0&0&0
    \end{pmatrix} \; ,\quad
    W^6=\begin{pmatrix}
        0&0&0&0&0\\
        0&0&1&0&0\\
        0&1&0&0&0\\
        0&0&0&0&0\\
        0&0&0&0&0
    \end{pmatrix} \; ,\\
    W^7&=\begin{pmatrix}
        0&0&0&0&0\\
        0&0&-i&0&0\\
        0&i&0&0&0\\
        0&0&0&0&0\\
        0&0&0&0&0
    \end{pmatrix} \; ,\quad
    W^8=\begin{pmatrix}
        \frac{1}{\sqrt{3}}&0&0&0&0\\
        0&\frac{1}{\sqrt{3}}&0&0&0\\
        0&0&\frac{-2}{\sqrt{3}}&0&0\\
        0&0&0&0&0\\
        0&0&0&0&0
    \end{pmatrix} \; ,\quad
    W^9=\begin{pmatrix}
        0&0&0&1&0\\
        0&0&0&0&0\\
        0&0&0&0&0\\
        1&0&0&0&0\\
        0&0&0&0&0
    \end{pmatrix} \; ,\\
    W^{10}&=\begin{pmatrix}
        0&0&0&-i&0\\
        0&0&0&0&0\\
        0&0&0&0&0\\
        i&0&0&0&0\\
        0&0&0&0&0
    \end{pmatrix} \; ,\quad
    W^{11}=\begin{pmatrix}
        0&0&0&0&0\\
        0&0&0&1&0\\
        0&0&0&0&0\\
        0&1&0&0&0\\
        0&0&0&0&0
    \end{pmatrix} \; ,\quad
    W^{12}=\begin{pmatrix}
        0&0&0&0&0\\
        0&0&0&-i&0\\
        0&0&0&0&0\\
        0&i&0&0&0\\
        0&0&0&0&0
    \end{pmatrix} \; ,\\
    W^{13}&=\begin{pmatrix}
        0&0&0&0&0\\
        0&0&0&0&0\\
        0&0&0&1&0\\
        0&0&1&0&0\\
        0&0&0&0&0
    \end{pmatrix} \; ,\quad
    W^{14}=\begin{pmatrix}
        0&0&0&0&0\\
        0&0&0&0&0\\
        0&0&0&-i&0\\
        0&0&i&0&0\\
        0&0&0&0&0
    \end{pmatrix} \; ,\quad
    W^{15}=\begin{pmatrix}
        \frac{1}{\sqrt{6}}&0&0&0&0\\
        0&\frac{1}{\sqrt{6}}&0&0&0\\
        0&0&\frac{1}{\sqrt{6}}&0&0\\
        0&0&0&\frac{-3}{\sqrt{6}}&0\\
        0&0&0&0&0
    \end{pmatrix} \;, \\
    W^{16}&=\begin{pmatrix}
        0&0&0&0&1\\
        0&0&0&0&0\\
        0&0&0&0&0\\
        0&0&0&0&0\\
        1&0&0&0&0
    \end{pmatrix} \; ,\quad
    W^{17}=\begin{pmatrix}
        0&0&0&0&-i\\
        0&0&0&0&0\\
        0&0&0&0&0\\
        0&0&0&0&0\\
        i&0&0&0&0
    \end{pmatrix} \; ,\quad
    W^{18}=\begin{pmatrix}
        0&0&0&0&0\\
        0&0&0&0&1\\
        0&0&0&0&0\\
        0&0&0&0&0\\
        0&1&0&0&0
    \end{pmatrix} \; ,\\
    W^{19}&=\begin{pmatrix}
        0&0&0&0&0\\
        0&0&0&0&-i\\
        0&0&0&0&0\\
        0&0&0&0&0\\
        0&i&0&0&0
    \end{pmatrix} \; ,\quad
    W^{20}=\begin{pmatrix}
        0&0&0&0&0\\
        0&0&0&0&0\\
        0&0&0&0&1\\
        0&0&0&0&0\\
        0&0&1&0&0
    \end{pmatrix} \; ,\quad
    W^{21}=\begin{pmatrix}
        0&0&0&0&0\\
        0&0&0&0&0\\
        0&0&0&0&-i\\
        0&0&0&0&0\\
        0&0&i&0&0
    \end{pmatrix} \; ,\\
    W^{22}&=\begin{pmatrix}
        0&0&0&0&0\\
        0&0&0&0&0\\
        0&0&0&0&0\\
        0&0&0&0&1\\
        0&0&0&1&0
    \end{pmatrix} \; ,\quad
    W^{23}=\begin{pmatrix}
        0&0&0&0&0\\
        0&0&0&0&0\\
        0&0&0&0&0\\
        0&0&0&0&-i\\
        0&0&0&i&0
    \end{pmatrix} \; ,\quad
    W^{24}=\begin{pmatrix}
        \frac{1}{\sqrt{10}}&0&0&0&0\\
        0&\frac{1}{\sqrt{10}}&0&0&0\\
        0&0&\frac{1}{\sqrt{10}}&0&0\\
        0&0&0&\frac{1}{\sqrt{10}}&0\\
        0&0&0&0&\frac{-4}{\sqrt{10}}
    \end{pmatrix} \;.
\end{align}
\end{widetext}

\section{Additional three-band models with symmetry-induced EP3s}\label{app:C}

\begin{figure}
    \centering
    \includegraphics[width=\columnwidth]{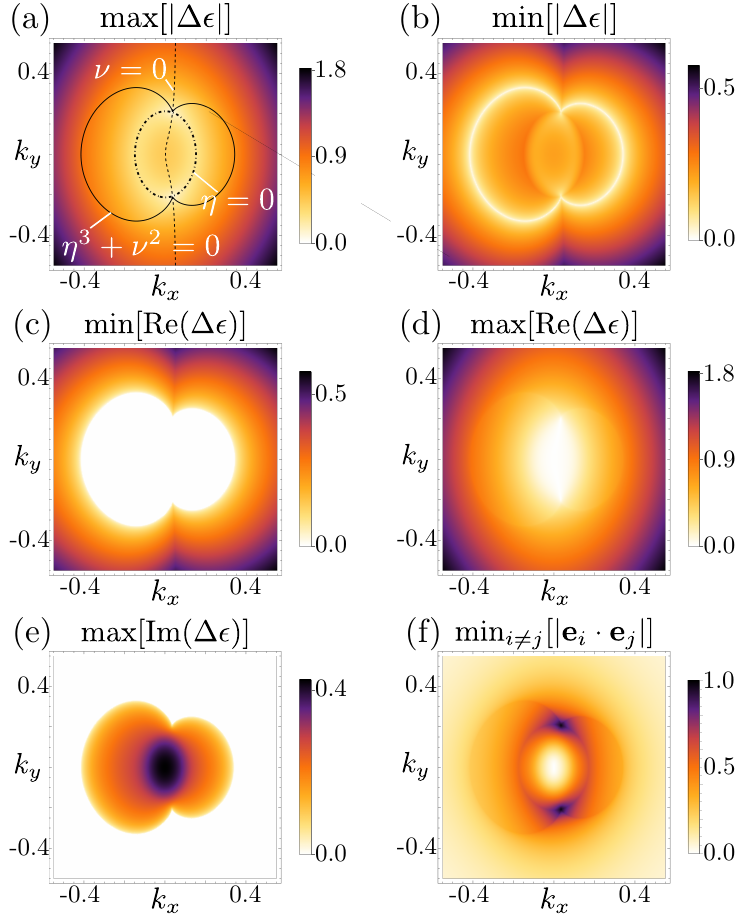}
    \caption{Spectral structure of psH-symmetric model defined by Eq.~(\ref{eq:psH}) with $\xi=0.15$: (a) shows the EP3 pair by plotting the largest value of the absolute value of the complex gap $\Delta \epsilon$. Further the curves given by the constraints are plotted; (b) highlights the EP2 lines connecting the EP3 pair by showing the minimum of $|\Delta \epsilon|$; (c) shows the two-level FS and (d) highlights the three-level FA separating the two two-level FSs; ; (e) emphasizes the three-level i-FS; (f) shows the the minimum overlap of any pair of eigenvectors to prove that the threefold degeneracies are truly EP3s.}
    \label{fig:psH}
\end{figure}

\emph{Model for psH-symmetry-induced EP3.}---The generic spectral structure in Fig.~\ref{fig:PT} derived in the main text can be shown for a Hamiltonian with psH symmetry in addition to the $\mathcal{PT}$-symmetric example provided in the main text.
The model Hamiltonian is given by
\begin{equation}
    \begin{split}
        H_\textrm{psH}(\bm{k})  = &\sin(k_x) \left(M^3+M^8\right) + \sin(k_y) M^5  \\ &+ h_s M^4 + i\xi \, \left(M^1+M^6\right) ,
    \end{split}\label{eq:psH}
\end{equation}
with $h_s=2-\cos(k_x)-\cos(k_y)$.
This model obeys psH symmetry $H_\textrm{psH}(\bm{k})=\varsigma H^*_\textrm{psH}(\bm{k})\varsigma^{-1}$ with the generator $\varsigma=\frac{\mathbb{1}_3}{3} + M^3 - \frac{M^8}{\sqrt{3}}$.
The spectral structure is presented in Fig.~\ref{fig:psH} and it shows the generic features we derived.

\begin{figure}
    \centering
    \includegraphics[width=\columnwidth]{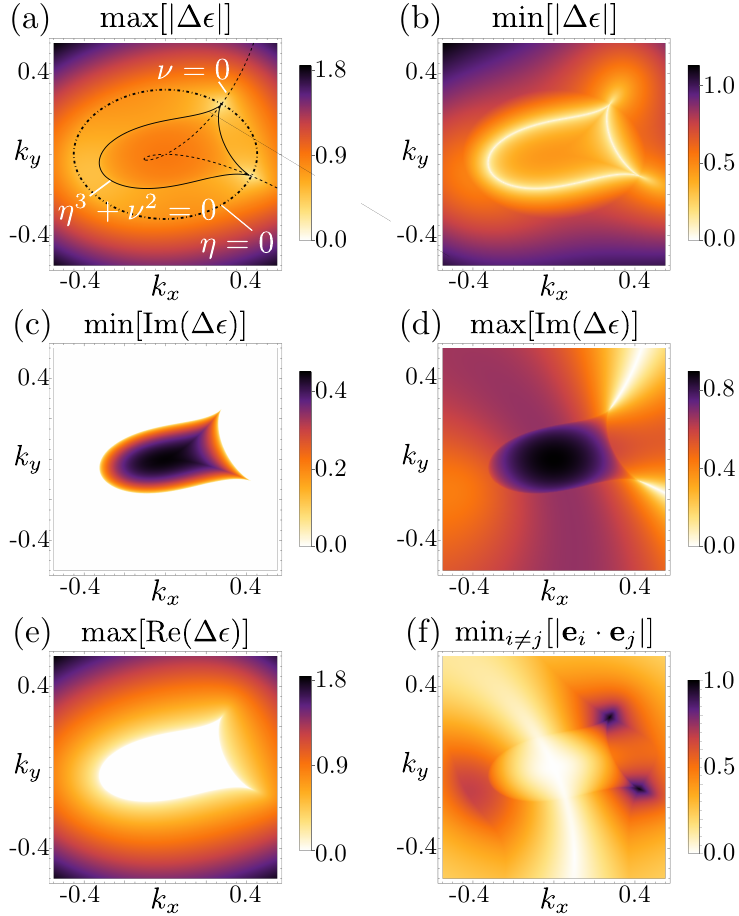}
    \caption{Spectral structure of CS model defined by Eq.~(\ref{eq:CS}) with $\xi=0.2$: (a) shows the EP3 pair by plotting the largest value of the absolute value of the complex gap $\Delta \epsilon$. Further the curves given by the constraints are plotted; (b) highlights the EP2 lines connecting the EP3 pair by showing the minimum of $|\Delta \epsilon|$; (c) shows the two-level i-FS and (d) highlights the three-level i-FA separating the two two-level FSs; (e) emphasizes the three-level FS; (f) shows the the minimum overlap of any pair of eigenvectors to prove that the threefold degeneracies are truly EP3s.}
    \label{fig:CS}
\end{figure}

\emph{Model for CS-induced EP3.}---The generic spectral structure in Fig.~\ref{fig:CP} can also be shown for a Hamiltonian with CS.
The model Hamiltonian is given by
\begin{equation}
    \begin{split}
        H_\textrm{CS}(\bm{k})  = &\sin(k_x) M^1 + \sin(k_y) \left(M^2+M^7\right)  \\ &+ h_s M^6 + i\xi \, \left(2\,M^4+M^5\right) ,
    \end{split}\label{eq:CS}
\end{equation}
with $h_s=2-\cos(k_x)-\cos(k_y)$.
This model obeys CS $H_\textrm{CS}(\bm{k})=-\Gamma H^\dagger_\textrm{CS}(\bm{k})\Gamma^{-1}$ with the generator $\Gamma=\frac{\mathbb{1}_3}{3} + M^3 - \frac{M^8}{\sqrt{3}}$.
The spectral structure is presented in Fig.~\ref{fig:CS} and it shows the generic features we derived.

\begin{figure}
    \centering
    \includegraphics[width=\columnwidth]{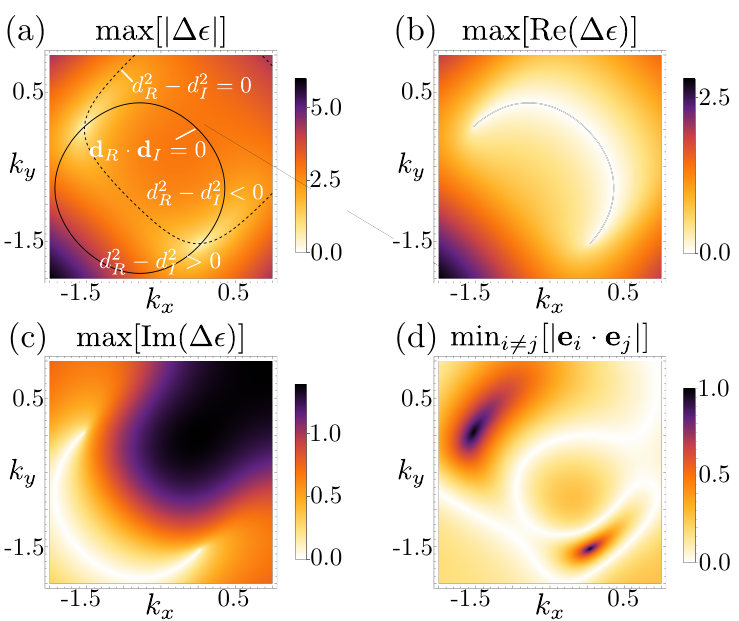}
    \caption{Spectral structure of SLS-induced EP3s for the model defined in Eq.~(\ref{eq:SLS}) with $\xi=0.4$. (a) highlights the EP3 pair, while also depicting the closed curves on which the constraints are fulfilled; (b) and (c) show the real and imaginary three-level Fermi arcs, respectively; (d) displays the minimum overlap of the eigenvector pairs of the model to verify the third order EPs.}
    \label{fig:SLS}
\end{figure}

\emph{Model for SLS-induced EP3.}---For a Hamiltonian with SLS the spectral structure derived in the main text can be shown in addition to the psCS example.
The model Hamiltonian is given by
\begin{equation}
    \begin{split}
        H_\textrm{SLS}(\bm{k})  = &\sin(k_x) M^1 + h_s M^6 + \sin(k_y) M^7\\  
        &+ i\xi \, \left(M^1+M^6+M^7\right) ,
    \end{split}\label{eq:SLS}
\end{equation}
with $h_s=2-\cos(k_x)-\cos(k_y)$.
This model obeys SLS $H_\textrm{SLS}(\bm{k})=\mathcal{S} H^*_\textrm{SLS}(\bm{k})\mathcal{S}^{-1}$ with the generator $\mathcal{S}=\frac{\mathbb{1}_3}{3} + M^3 - \frac{M^8}{\sqrt{3}}$.
The spectral structure is presented in Fig.~\ref{fig:SLS}.
We observe the EP3 pair and the two three-level Fermi arcs connecting them exactly as expected.

\section{Possibility of indistinguishable EP2s in presence of psCS and SLS}\label{app:D}

\begin{figure}[t]
    \centering
    \includegraphics[width=\columnwidth]{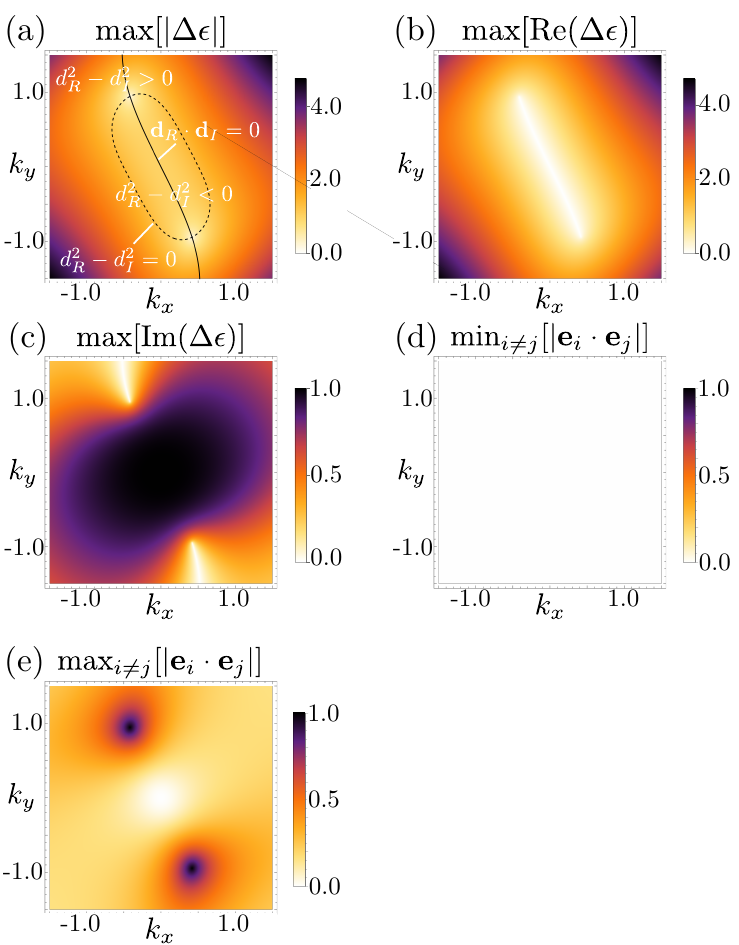}
    \caption{Spectral structure of EP2s on top of orthogonal flat band in the presence of psCS for the model presented in Eq.~(\ref{eq:EP2}) with $\xi=0.5$. (a) highlights the EP2 pair, while depicting the closed curves on which the constraints are fulfilled; (b) and (c) show the real and imaginary three-level FSs, respectively; (d) displays the minimum overlap of the eigenvector pairs, which is always 0, since the flat band is orthogonal to the EP2 structure, and (e) shows the maximum overlap of eigenvector pairs to prove that the degeneracies are in fact EP2s.}
    \label{fig:EP2}
\end{figure}

In the presence of psCS and SLS, the EP3s appearing in a three-band system look identical to a situation in which EP2s appear together with a flat band. As such, there is a question on how to distinguish these two cases. Here, we show that EP2s cannot appear in a three-band system with psCS, whereas we present an example for EP2s in a three-band model with SLS.

\emph{Proof of the impossibility of EP2s in three-band systems with psCS.}---Let us show that psCS prohibits the appearance of an EP2 plus an orthogonal band in three-band systems.
Due to the symmetry the EP2s would have to occur on top of the flat band at zero energy, i.e., they would have a three-fold degeneracy in the eigenvalue spectrum but only two eigenvectors coalesce onto one.
If such a system would exist we would be able to write it as
\begin{equation} \label{eq:EPsplus1}
    H_\textrm{EP2+1} = \begin{pmatrix}
        H_\textrm{EP2} & \begin{matrix}
            0 \\
            0
        \end{matrix} \\
        \begin{matrix}
            0 & 0
        \end{matrix} & \epsilon_0
    \end{pmatrix} ,
\end{equation}
where $\epsilon_0=0$ is the energy of the flat band, and $H_\textrm{EP2}$ is traceless.
If we enforce psCS on this system the symmetry generator has to have the form
\begin{equation}
    X_3 = \begin{pmatrix}
        X_2 & \begin{matrix}
            0 \\
            0
        \end{matrix} \\
        \begin{matrix}
            0 & 0
        \end{matrix} & 1
    \end{pmatrix},
\end{equation}
with $X_2$ being the generator of psCS of the two-band Hamiltonian $H_\textrm{EP2}$, i.e., $H_\textrm{EP2}^T ({\bf k}) = -X_2 H_\textrm{EP2} ({\bf k})X_2^{-1}$ with $X_2^2 = 1$.
The physical constraints on the system are independent of the choice of operator so we can choose any of the three Pauli matrices for $X_2$.
Writing $H_\textrm{EP2} = \bm{d}(\bm{k}) \cdot \bm{\sigma}$ with $\bm{\sigma}$ the vector of Pauli matrices, and choosing $X_2=\sigma_z$, we find psCS yields $d_y=d_z=0$~\cite{Sayyad2022}.
The only remaining term in $H_\textrm{EP2}$ is $d_x=d_{x,R}+id_{x,I}$, such that the eigenvalue read $\pm \sqrt{d_{x,R}^2 - d_{x,I}^2 + 2 i d_{x,R}d_{x,I}} $. A two-fold degeneracy can thus only be found iff $d_{x,R} = d_{x,I} = 0$, which amounts to finding an ordinary degeneracy. In other words, in the presence of psCS it is not possible to realize EP2s in a three-band system, and the threefold degeneracies must thus always correspond to EP3s. 

\emph{Model for an EP2 in a three-band system with SLS.}---Let us start by first using the same line of reasoning as for the psCS case. To show that one can find an EP2 in a three-band system with SLS, we start with the Hamiltonian in Eq.~\eqref{eq:EPsplus1}. To enforce SLS, the symmetry generator reads
\begin{equation}
    \mathcal{S}_3 = \begin{pmatrix}
        \mathcal{S}_2 & \begin{matrix}
            0 \\
            0
        \end{matrix} \\
        \begin{matrix}
            0 & 0
        \end{matrix} & 1
    \end{pmatrix},
\end{equation}
where $\mathcal{S}_2$ is the generator of SLS in the two-band Hamiltonian $H_\textrm{EP2}$, i.e., $H_\textrm{EP2} ({\bf k}) = -\mathcal{S}_2 H_\textrm{EP2}({\bf k})\mathcal{S}_2^{-1}$ and $\mathcal{S}_2^2 =1$. Writing $H_\textrm{EP2}$ in terms of the Pauli matrices, we can choose any of the Pauli matrices for $\mathcal{S}_2$, such that choosing $\mathcal{S}_2 = \sigma_z$ yields $d_z = 0$ and $H_\textrm{EP2} = d_x \sigma_x + d_y \sigma_y$~\cite{Sayyad2022}. Therefore, it is possible to find EP2s in a three-band system with SLS.

Let us present an example in which EP2s appear in a three-band system. The Hamiltonian reads
\begin{equation}
    \begin{split}
        H_{\textrm{EP2}+1}(\bm{k})  = &\left[\sin(k_x)+\frac{1}{2}\sin(k_y)\right]M^1+ \\
        &h_s M^2 +i\xi M^1,
    \end{split}\label{eq:EP2}
\end{equation}
with $h_s=2-\cos(k_x)-\cos(k_y)$.
Here, the generator of SLS $H_{\textrm{EP2}+1}(\bm{k})=-\mathcal{S}_2 H_{\textrm{EP2}+1}(\bm{k})\mathcal{S}_2^{-1}$ reads $\mathcal{S}_2=\frac{\mathbb{1}_3}{3} + M^3 - \frac{M^8}{\sqrt{3}}$.
In this model only the first two Gell-Mann matrices contribute and thus the third band is isolated.
In Fig.~\ref{fig:EP2} the spectral structure is shown and it is indistinguishable from the structure of EP3s induced by SLS. What is more, the constraints for finding EP2s, cf. Fig.~\ref{fig:EP2}(a), are identical to the constraints for finding EP3s, cf. Fig.~\ref{fig:psCS}(a).

\end{document}